\documentclass[twocolumn,showpacs,preprintnumbers,amsmath,amssymb]{revtex4}
\usepackage{graphicx}
\usepackage{dcolumn}
\usepackage{bm}

\usepackage{epsfig}
\begin{document}
\title{Theory of Single File Diffusion in a Force Field 
}
\author{E. Barkai$^{1,2}$}
\author{R. Silbey$^1$}
\affiliation{
$^1$Department of Chemistry, Massachusetts Institute of Technology,
Cambridge, Massachusetts 02139, USA \\
$^2$ Department of Physics,
Bar Ilan University,
Ramat-Gan 52900 Israel}
\begin{abstract}
{
The dynamics of  hard-core interacting
 Brownian particles in an external 
potential field is studied in one dimension. 
Using the Jepsen line we find a very general and
 simple formula relating the motion
of the tagged center particle, 
with the classical, time dependent single particle
reflection ${\cal R}$ and transmission ${\cal T}$
coefficients.
Our
formula describes rich physical behaviors both in equilibrium 
and the approach to equilibrium of this many body problem.  
}
\end{abstract}

\maketitle

 Single file diffusion of a tagged Brownian particle, 
interacting 
with other Brownian particles, is a model for motion of
a molecule or particle in a crowded one-dimensional environment, such as
a biological pore or channel
\cite{Hanggi,Eok} and for experimentally studied physical systems
such as zeolites \cite{Hahn} and confined colloid particles \cite{Lutz,Haim}.
The confinement of the tagged particle by the other particles is strong
and severely restricts the motion of the particle. The description of
this single file motion has been of much theoretical interest
\cite{Hanggi,Harris,Levitt,Kollmann,Henk,Roden,Taloni,Tobias,Aslangul,Flomenbom}.
For an infinite system and uniform initial particle density,
Harris \cite{Harris} and Levitt \cite{Levitt} first showed that a 
tagged particle exhibits anomalous diffusion
 $\langle (x_T)^2 \rangle \sim t^{1/2}$, 
rather then normal diffusion $\langle (x_T)^2 \rangle \sim t$, 
due to the strong many body confinement effect. In recent years, 
two important research direction have begun to emerge. First, 
the effect of an external force field acting on the particles is important
since, in many cases, pores induce entropic barriers \cite{Hanggi}
and are generally
inhomogeneous; hence, single file motion in a periodic potential
\cite{Taloni} and a confining box \cite{Tobias} were investigated.
Secondly, initial conditions have a profound effect on single file
motion \cite{Aslangul,Flomenbom}: for example, particles with initial
delta function distribution in space (rather than a uniform 
distribution as assumed in \cite{Harris,Levitt})
 yield normal diffusion \cite{Aslangul}.
This is important since if a potential field is acting on the particles, 
thermal initial conditions will have a Boltzmann weighting, leading to generally
non-uniform initial conditions.

Here we provide a general and surprisingly simple
theory  of single file diffusion 
 valid in the presence (or absence) of
a potential field, $V(x)$, as well as  for thermal and
non-thermal
initial conditions. In addition,  we briefly discuss
why our main results are valid even for 
the case where the underlying dynamics is anomalous
 \cite{remark1}. 
Our general  result reproduces those previously obtained
as well as many new ones, 
by mapping the many particle problem onto a solvable single particle model. 

 In our model, 
$2 N +1$ identical 
particles with hard core particle-particle interactions
 are undergoing Brownian
motion in one dimension, so particles cannot pass one another.  
An external potential field $V(x)$ is acting on the particles.
We tag the central particle, which has $N$ other particles to its left,
and $N$ to its right.
Initially the  tagged particle is at $x=0$. 
The motion of a single particle, in the absence of
interactions with other particles,
 is over-damped Brownian motion
so that the single non interacting particle Green function
$g(x,x_0,t)$, with the initial condition $g(x,x_0,0) = \delta(x-x_0)$,
is obtained from the Fokker-Planck equation \cite{Risken}
\begin{equation}
{\partial g(x,x_0,t) \over \partial t} = 
 D \left[ {\partial^2 \over \partial x^2} - { 1 \over k_b T} {\partial \over \partial x} F\left(x \right) \right] g(x,x_0,t),
\label{eq000}
\end{equation}
and $F(x) = - V'(x)$ is the force field.

\begin{figure}[!htb]
\begin{center}
\includegraphics[width=5.cm,height=8cm,angle=-90]{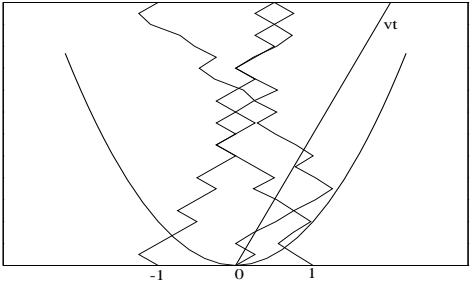}%
\caption{\label{fig1} Schematic motion of Brownian particles in a 
harmonic potential (the parabola) where particles cannot
penetrate through each other. 
The straight line is called the Jepsen line, as explained in the text. 
The center tagged particle is labeled $0$. 
In an equivalent non-interacting picture,
 we allow particles to pass through each other,
and we follow the trajectory of the particle which is at the center.
}
\end{center}
\end{figure}

In Fig. \ref{fig1} a schematic diagram of the problem is presented. 
  The straight line is called the Jepsen line, which starts from the
origin $x=0$
and follows the rule $x(t)=v t$, where $v$ is a test velocity
\cite{Levitt}. 
In the interacting system we label particles according to their
initial position increasing to the right (see Fig. 1). 
As noticed in \cite{Jepsen,Levitt},
since the Brownian particles are impermeable, every time a particle
crosses the Jepsen line from the right (or left), 
the particle number immediately
to the left of the line will be raised (or lowered) by one, respectively.  
Hence, the particle number immediately to the left of the Jepsen line
defines a stochastic process
decreasing and increasing its value  $+1$ or $-1$
or zero randomly. 

 Following Levitt \cite{Levitt} we now consider a non interacting system,
equivalent to
the interacting one in the large $N$ limit. 
We let particles pass through each other, 
but switch labels upon collision, 
and introduce the 
counter $\alpha(t)$ which increases by $+1$ if a particle
crosses the Jepsen line from left, and decreases by $-1$ when
a particle crosses this line from right.  
The event when the counter $\alpha$ 
has the value zero, is equivalent to finding the tagged particle
to the left of the Jepsen line in the interacting system.
This is the case since then
the total number of crossings from left to right is equal
the number of crossings from right to left.
So the probability
of the  tagged particle
being in the vicinity of $x_T=v t$ is given by the probability that
$\alpha=0$, i.e. by the statistics of the number of transitions 
of the Jepsen line \cite{Levitt}. In what follows we depart from
the approach in \cite{Levitt,remark3}. 

Our aim is to calculate the probability of the random variable
$\alpha$,  $P_N(\alpha)$, and then switch $v t \to x_T$ to find the
probability density function (PDF) of the tagged particle. 
For that we designate
 $P_{LL}(x_0 ^{-j})$ as the probability that a non tagged particle $j$
starting  to the left of the Jepsen line $x_0 ^{-j}<0$, 
is found also at time $t$ on the left 
of this line.
 $P_{LR}$ is  the probability of a particle starting to
the left of the Jepsen line to 
end on the right, and similarly for $P_{RR}$ and  $P_{RL}$. 
 Consider first
 $N=1$, that is, one particle which starts at $x_0 ^{1}>0$ and
a second particle which starts at $x_0 ^{-1}<0$. 
Then clearly we have
either $\alpha=\pm 1$ or $\alpha=0$. The probabilities of these
events are easily calculated, for example
$P_{N=1} \left( \alpha=1 \right) = 
  P_{LR} (x_0 ^{-1}) P_{RR} (x_0 ^{+1})$  
is the probability that one particle crossed from $L$ to $R$
and the other remained in domain $R$. 
 Similarly 
  $P_{N=1} \left( \alpha=0 \right)= P_{LL} (x_0 ^{-1}) P_{RR} (x_0 ^{+1})  
 +   P_{LR} (x_0 ^{-1}) P_{RL} (x_0 ^{+1}) $ and 
 $P_{N=1} (\alpha=-1)=   P_{LL} (x_0 ^{-1}) P_{RL} (x_0 ^{+1})$.  
Since in the non-interacting picture, the motion of the particles
are independent, we can use random walk theory and
 Fourier analysis \cite{Montroll} to find 
the behavior for any $N$
\begin{equation}
P_N \left(\alpha\right) = { 1 \over 2 \pi} \int_{-\pi} ^\pi {\rm d} \phi \Pi_{j=1} ^N 
\lambda\left( \phi, x_0 ^{-j} , x_0 ^{j} \right) e^{ i \alpha \phi}.
\label{eq02}
\end{equation}
Where the structure function is
$$ \lambda\left(\phi,x_0 ^{-j}, x_0 ^{j} \right) = 
e^{i \phi} P_{LR} (x_0 ^{-j}) P_{RR} (x_0 ^{j}) + 
P_{LL}(x_0 ^{-j}) P_{RR} (x_0 ^{j}) $$
\begin{equation}
+
P_{LR}(x_0 ^{-j}) P_{RL} (x_0 ^{j})   + e^{-i \phi}
P_{LL}(x_0 ^{-j}) P_{RL} (x_0 ^{j}).  
\label{eq03}
\end{equation}
We average Eq. (\ref{eq02}) with respect to the initial conditions
$x_0^{j}$ and $x_0 ^{-j}$, which are assumed to be independent identically 
distributed
random variables and we find
\begin{equation}
\langle P_N \left( \alpha \right) \rangle=  
 { 1 \over 2 \pi} \int_{-\pi} ^\pi {\rm d} \phi \langle \lambda\left(\phi\right) \rangle^N e^{ i \alpha \phi} 
\label{eq04}
\end{equation} 
where
from Eq. (\ref{eq03})
the averaged structure function is
$$ \langle \lambda\left(\phi\right) \rangle =  $$
\begin{equation}
e^{i \phi} \langle P_{LR} \rangle \langle  P_{RR} \rangle + 
\langle P_{LL}\rangle  \langle P_{RR} \rangle + 
\langle P_{LR} \rangle \langle P_{RL} \rangle   + e^{-i \phi}
\langle P_{LL} \rangle \langle  P_{RL} \rangle.  
\label{eq05}
\end{equation} 
The averages in Eq. (\ref{eq05})
are easy to find in principle, in terms of the Green function of
the non-interacting particle and the initial density of particles,
for example
\begin{equation}
\langle P_{LR} \rangle = 
\int_{-\overline{L}} ^0 f_{L}(x_0) \int_{x_T} ^{\overline{L}}
 g(x,x_0,t) {\rm d} x {\rm d} x_0
\label{eq09a}
\end{equation}
where the mentioned replacement $v t \to x_T$ was made.
Here $f_L(x_0)$ is the PDF
of initial positions of the particles which initially are at
$x_0<0$, similarly $f_R(x_0)$ describes the initial conditions of the
right particles. 
 In Eq. (\ref{eq09a})
$\overline{L}$ 
is the system size, 
which can be taken to 
infinity in the usual way.

Eq. (\ref{eq04}) describes a random walk where the number of particles
$N$ serves as an operational time. We may use the Gaussian central limit 
theorem (CLT) to analyze this walk, when $N \to \infty$. In that limit 
the first two moments of the structure function, $\mu_1$ and $\mu_2$,
found
in the small $\phi$ expansion 
\begin{equation}
\langle \lambda(\phi) \rangle = 1 + i \mu_1 \phi - {1 \over 2 } \mu_2 \phi^2 + O(\phi^3)
\label{eq06}
\end{equation}
are the only two parameters needed to determine the behavior of $P_N (\alpha)$.
  Defining the variance 
$\sigma^2 = \mu_2 - (\mu_1)^2$, using Eq. (\ref{eq05}) 
and the normalization condition e.g. $\langle P_{LR} \rangle + \langle P_{LL} 
\rangle = 1$ we find the expected result
\begin{equation}
\mu_1 = \langle P_{LR} \rangle - \langle P_{RL} \rangle 
\label{eq07}
\end{equation}
\begin{equation} 
\sigma^2 = 
  \langle P_{RR} \rangle \langle P_{RL} \rangle  
+ \langle P_{LL} \rangle \langle P_{LR} \rangle. 
\label{eq08}
\end{equation}
Using the CLT we have the probability of zero crossing, namely
$\alpha =0$ in the $N \to \infty$ limit
\begin{equation}
P_N\left( \alpha=0 \right) \sim { \exp\left(  - {N (\mu_1)^2 \over 2 \sigma^2} \right) \over \sqrt{ 2 \pi N } \sigma}.
\label{eq09}
\end{equation}
This is our first general result, valid for a large 
class of 
Green functions
and initial conditions and  thus suited for 
the investigation of a wide range of problems.

 Symmetric potential fields
$V(x)= V(-x)$, and symmetric initial conditions are now investigated.
The latter simply means that the
 density of the initial positions of the left
 particles, i.e. those residing initially in $x_0<0$,
is the same as that of  the right particles,
$f_R(x_0)= f_L (-x_0)$.
In this case the subscript $R$ and $L$ is redundant
and we use $f(x_0)=f_R(x_0)=f_L(-x_0)$ to describe the initial conditions \cite{remark4}.
 From symmetry it is clear that
the tagged particle is unbiased, namely $\langle x_T \rangle=0$.     
Further, since $N$ is large we may expand expressions in the $\exp$ in
Eq. (\ref{eq09}) in
$x_T$, to obtain leading terms 
\begin{equation}
\mu_1 = \Delta J x_T + O(x_T)^2 
\label{eq10}
\end{equation}
where   
we used the symmetry of the problem which
implies 
$(\langle P_{LR}\rangle - \langle P_{RL}\rangle)_{x_T =0} = 0$,
 and by definition
\begin{equation}
\Delta J = {\partial \over \partial x_T} 
\left[ \langle P_{LR}(x_T) \rangle - \langle P_{RL}\left( x_T\right)  \rangle \right]|_{x_T=0}.
\label{eqx}
\end{equation}
Similarly 
\begin{equation}
\sigma^2 =
2\langle P_{RR} (x_T) \rangle 
\left[ 1 - \langle  P_{RR} \left(x_T\right) \rangle \right]|_{x_T = 0}+O(x_T).
\label{eq12}
\end{equation}
We designate $\langle P_{RR}(x_T) \rangle|_{x_T=0} ={\cal R}$ as 
  a reflection coefficient,
since it is the probability that a particle starting at $x_0<0$ 
is found at $x<0$ at time $t$ when an average over all initial
conditions is made 
\begin{equation}
{\cal R}= \int_0 ^{\overline{L}} f(x_0) \int_0 ^{\overline{L}} g(x,x_0,t) {\rm d} x {\rm d} x_0. 
\label{eq13}
\end{equation}
As usual the transmission coefficient ${\cal T} =1- {\cal R}$ 
is defined through ${\cal T} = \langle P_{RL}(x_T) \rangle|_{x_T=0}$. 
Notice that these reflection and transmission coefficients are
time dependent  single particle
quantities which give useful information for the 
many body problem. 
Also from symmetry we have 
$ {\partial \over \partial x_T} \langle P_{RL}(x_T) \rangle |_{x_T=0}
=-{\partial \over \partial x_T} \langle P_{LR}( x_T) \rangle |_{x_T=0}$
in Eq. (\ref{eqx}).
Hence we define
$j=-\partial \langle P_{LR} (x_T) \rangle / \partial x_T | _{x_T =0}$ where
from its definition Eq.  
(\ref{eq09a})
\begin{equation}
j= \int_0 ^{\overline{L}} f(x_0) g(0,x_0,t) {\rm d} x_0.
\label{eq14}
\end{equation}
So $j$ is the
density of non-interacting particles at $x=0$ for an initial density $f(x_0)$. 
Using Eqs. 
(\ref{eq09}-\ref{eq12})
we find our main result the PDF of the tagged particle
\begin{equation}
P(x_T) \sim  {1 \over \sqrt{ 2 \pi \langle \left( x_T \right)^2 \rangle}}
\exp\left[ - { \left( x_T \right)^2 \over 2 \langle \left( x_T \right)^2 \rangle} \right]
\label{eq15}
\end{equation}
where 
$\langle \left( x_T \right)^2 \rangle = 
{\cal R} {\cal T} / ( 2 N j^2)$ 
is the mean square displacement (MSD).

{\em Gaussian packet}. As our first example, we consider particles
without external forces $V(x)=0$,
in an infinite system with symmetric
Gaussian
initial conditions with a width $\xi$: 
$f(x_0) = \sqrt{2} \exp( - x_0 ^2 / (2 \xi^2) ))/ \sqrt{\pi \xi^2}$
$(x_0>0)$. 
The free particle Green function is 
\begin{equation}
g(x,x_0,t)={\exp\left( - {(x - x_0)^2 \over 4 D t } \right) \over \sqrt{ 4 \pi  D t}}.
\label{eq15dd}
\end{equation}
Using Eqs. (\ref{eq13}-\ref{eq15dd}) we find
the MSD of the tagged particle \cite{remark}
\begin{equation}
\langle (x_T)^2\rangle  \sim \xi^2  {\pi \over N} \left( 1 + { 2 D t \over \xi^2}  \right) \left[ { 1 \over 4} - { 1 \over \pi^2} \mbox{arccot}^2 \left( \sqrt{ {2 D t \over \xi^2}  } \right) \right].  
\label{eq15cc}
\end{equation}
This solution exhibits a transition from anomalous sub-diffusion 
to normal diffusion. At short times $2 D t/\xi^2 \ll 1$ 
%
$\langle (x_T)^2 \rangle \sim \xi {\sqrt{ 2 D t} \over N}$
%
while at long times
$\langle (x_T)^2 \rangle \sim {\pi D t \over 2 N}$.
%
For short times the particles do not have time to disperse, hence
the motion of the tagged particle is slower than normal, increasing
as $t^{1/2}$ since it is confined by the other particles in
the system.

{\em Particles in a box}. We  now consider the example of
 particles in a box extending
from $- \overline{L}$ to $\overline{L}$,
 which was recently investigated using the Bethe--ansatz and
numerical simulations \cite{Tobias}.
The tagged particle initially on $x=0$ has $N$ particles
uniformly distributed to its left and similarly to its right. In this case
$f(x_0) = 1/ \overline{L}$ for $0<x_0<\overline{L}$. 
The green function $g(x,x_0,t)$
 of a single particle in a box with
reflecting walls is found using an eigenvalue approach 
\cite{Risken}, similar to the method of solution
of the 
undergraduate quantum mechanical problem of a particle in a box.  
We find  
\begin{equation}
{\cal R} = {1\over 2} + { 4 \over \pi^2} \sum_{n=1, \mbox{Odd}} ^\infty
{ \exp \left[ - D {\left( \pi^2 n^2\right) \over 4 \overline{L}^2} t \right] \over
n^2} 
\label{eq16}
\end{equation}
so at $t=0$, ${\cal R} = 1$ since all particles initially say in
$(0,\overline{L})$ did
not have time to move to the other side of box, and 
$\lim_{t \to \infty} {\cal R} =1/2$
since in the long time limit there is equal probability for
a non-interacting particle to occupy half of the box.
The eigenvalues of the non-interacting particle
control the {\em exponential}
decay of ${\cal R}$  Eq. (\ref{eq16}) which in turn
determines the dynamics of the interacting tagged particle. 
We also find $j=1/2 \overline{L}$ using Eq. (\ref{eq14}).
For short times $D \pi^2 t / 4 \overline{L}^2 \ll 1$
we can replace the
summation in Eq. (\ref{eq16}) with integration and then using Eq. (\ref{eq15})
and $\rho = N / \overline{L}$  
\begin{equation}
P\left(x_T \right) \sim {\sqrt{\rho} \over 2 \left( \pi D t \right)^{1/4}} 
\exp\left[ - {\rho \left( x_T \right)^2 \sqrt{\pi} \over 4 \sqrt{ D t } } \right].
\label{eq17}
\end{equation}
Thus the tagged particle undergoes a single file
 sub-diffusive process \cite{Harris,Levitt}
$\langle (x_T)^2 \rangle \sim  2 (D t)^{1/2}/ (\rho \sqrt{\pi})$,
since for short times the particles do not interact with walls. 
In the long time limit, the tagged particle reaches an
equilibrium easily found using Eqs. (\ref{eq15},\ref{eq16})
\begin{equation}
p\left( x_T \right) \sim { \sqrt{  N } \over \sqrt{ \pi} \overline{L} } e^{ - { N (x_T)^2 \over \overline{L}^2} }
\label{eq18}
\end{equation}
and $\lim_{t \to \infty} \langle (x_t)^2 \rangle = \overline{L}^2/ (2 N)$.
More generally from Eq.  
(\ref{eq15}),
$\langle(x_T)^2 \rangle=2{\cal R}\left( 1 - {\cal R} \right)\overline{L}^2/N$
which nicely matches and simplifies considerably,
the Bethe-ansatz solution \cite{Tobias}
already for $N=70$ \cite{remark5}.

{\em Thermal initial conditions}. If we assume that initially the particles
are in thermal equilibrium, our simple formulas simplify even more.
If $f(x_0) = 2 \exp( - V(x_0)/ k_b T)/ Z$ where 
the normalizing partition function is 
\begin{equation}
Z = \int_{-\infty} ^\infty \exp \left[ - {V\left(x\right) \over k_b T} \right] 
{\rm d} x
\label{eqZ}
\end{equation}
then $j=1/Z$.
To see this, note that since we have a symmetric case $V(x) = V(-x)$ then
$g(0,-x_0,t) = g( 0, x_0,t)$ and 
$j = 
2 \int_0 ^\infty \exp\left[ - V(x_0) / k_b T \right]  g(0,x_0,t) {\rm d} x_0/Z$
gives
\begin{equation}
j= \int_{-\infty} ^\infty {\exp\left[ -{ V\left( x_0 \right) \over k_b T } \right] \over Z} g(0,x_0, t) {\rm d} x_0.
\label{eqjj}
\end{equation}    
Where we assume that the potential is binding so a stationary solution of
the Fokker-Planck equation is reached i.e. free particle  Eq. 
(\ref{eq15dd}) is excluded. 
Therefore $j$ Eq. (\ref{eqjj})
is the probability of finding non-interacting particles on the origin,
with thermal equilibrium  initial conditions. 
Since the latter is  the
stationary solution of the Fokker-Planck operator,
$j$ is time independent  and
equal to $j= \exp[ - V(0)/ k_b T]/Z$.
We can always choose $V(0)= 0$ and then $j = 1/Z$. 
Using Eq. 
(\ref{eq15})
we find the MSD of the tagged particle 
\begin{equation}
\langle (x_T)^2 \rangle ={ {\cal R} {\cal T} Z^2\over  2 N}.
\label{eqsimple}
\end{equation} 

{\em Single file motion in Harmonic Potential.} 
We now consider particles
in an Harmonic potential field 
$V(x) = m \omega^2 x^2 / 2$
where $\omega$ is the Harmonic frequency and
initially the particles
are in thermal equilibrium. 
The corresponding single particle green function $g(x,x_0,t)$ 
describes the Ornstein--Uhlenbeck process, and is well
known \cite{Risken}.
 Defining the thermal length scale
$\xi_{{\rm th}} = \sqrt{ k_b T / m \omega^2}$
and using  thermal initial conditions 
$f(x_0)=\sqrt{2 / \pi \xi_{{\rm th}} ^2}\exp \left[-(x_0)^2/ 2\xi_{{\rm th}} ^2 \right]$, 
we find using
Eqs. (\ref{eq13},\ref{eqZ},\ref{eqsimple})  
\begin{equation}
\langle \left( x_T \right)^2 \rangle = { \pi \over N} \left( \xi_{{\rm th}}\right)^2 \left\{  { 1 \over 4} - { 1 \over \pi^2} \mbox{arccot}^2 \left[ \sqrt{ \exp\left( 2 \tilde{t} \right) - 1} \right] \right\}  
\label{eqHO}
\end{equation}
where $\tilde{t} = D t / (\xi_{{\rm th}})^2$ is dimensionless time. 
For short times $\tilde{t} \ll 1$ we obtain sub-diffusive behavior
$\langle (x_T)^2 \rangle \sim \xi_{{\rm th}} \sqrt{ 2 D t} / N$ since 
then the effects of the binding field are negligible, while the tagged particle
motion is restricted by all others leading to sub-diffusive behavior. 
For long times the tagged particle reaches an equilibrium
$ \langle (x_{T})^2 \rangle \sim  \pi (\xi_{{\rm th}} )^2  / 4 N$.

{\em Equilibrium of tagged particle.} In the long time limit,
 and for binding potential fields we find again a simple limiting behavior.
 First, note that $\lim_{t \to \infty} {\cal R} = 1/2$
as is easily obtained from Eq. (\ref{eq13}) and physically
obvious for the symmetric system under investigation.  
Secondly, for any initial condition $g(0,x_0,t)= 1/Z$ when $t \to \infty$,
namely the Green function reached an equilibrium 
and hence we find from Eq. (\ref{eq14})
 $\lim_{t \to \infty} j = 1 / Z$. Therefore in the long time
limit 
\begin{equation}
\lim_{t \to \infty} \langle \left( x_T \right)^2 \rangle \sim 
{Z^2 \over 8 N}. 
\label{eqZN}
\end{equation}
Consistently, this result can be derived directly from the canonical ensemble,
using the many body Hamiltonian of the  system, 
in the large $N$ limit, without resorting to dynamics as we did here. 
The long time limit of our examples 
of  particles in a box
Eq. (\ref{eq18})
and in a harmonic field 
Eq. (\ref{eqHO})
are
easily recovered from the more general Eq. (\ref{eqZN}).

 In the examples analyzed so far,
we considered symmetric external fields, and initial 
conditions, which yield no drift $\langle x_T \rangle=0$.
Our general results Eqs. (\ref{eq09},\ref{eq15}) can be  extended to 
cases where the drift is non-zero, 
then expansions Eqs. (\ref{eq10},\ref{eqx},\ref{eq12})
include additional terms which make
the final expression slightly more cumbersome and 
hence suited for a longer publication.
 Another interesting extension of this work is the
 case where the particle dynamics,
 in the {\em absence} of interactions with each other, 
 is anomalous as was considered in \cite{Flomenbom,Band} for the
case of sub-diffusing particles \cite{remark1}. 
These results will be addressed in 
the future.

 To conclude, we have mapped the many body problem of interacting
hard core Brownian particles to a single particle problem
where calculation of the reflection coefficient ${\cal R}$ and
$j$ yield the motion of the tagged particle. In the large $N$ limit,
information
on the motion of the tagged particle is contained in the single particle
 Green function
$g(x,x_0,t)$ which can be calculated with known methods.
Rich physical behaviors emerge, which depend on the
initial distribution of the particles and the force field. 
For particles in a confining field, a transition from sub-diffusion
behavior to an equilibrium value is found to be controlled by
the eigen modes of the single particle green function, thus
the approach to equilibrium is fast in this sense, even though
the diffusion is slower than normal. In contrast, for Gaussian
initial conditions a transition between sub-diffusive to normal
diffusion is found, which is the expected transition for a wide variety
of well behaved initial conditions in the absence of a force.
The most rewarding part of this work
are the simple equations we find for the single  file problem.

{\bf Acknowledgments} EB is supported by the Israel Science foundation.
We thank
Tobias Ambj\"ornsson    
for
helpful discussions.


\begin{thebibliography}{99}

\bibitem{Hanggi} P.S. Burada, P. H\"anggi, F. Marchesoni
G. Schmid, and P. Talkner 
	arXiv:0808.2345v1 [cond-mat.stat-mech] (2008).

\bibitem{Eok} H. Kim, C. Kim, E. K. Lee, P. Talkner, and P. H\"anggi
{\em Phys. Rev. E} {\bf 77}, 031202 (2008). 


\bibitem{Hahn} K. Hahn, J. K\"argeri, and V. Kukla {\em Phys. Rev. Lett.} 
{\bf 76} 2762 (1996).

\bibitem{Lutz} C. Lutz, M. Kollmann, and C. Bechinger
{\em Phys. Rev. Lett.} {\bf 93} 026001 (2004).

\bibitem{Haim} B. Lin, M. Meron, B. Cui, S. A. Rice, H. Diamant
{\em Phys. Rev. Lett.} {\bf 94} 216001 (2005).

\bibitem{Harris} T. E. Harris {\em J. Appl. Prob.} {\bf 2}, 323 (1965).

\bibitem{Levitt} D. G. Levitt {\em Phys. Rev. A} {\bf 8} 3050 (1973).


\bibitem{Kollmann} M. Kollmann {\em Phys. Rev. Lett.} {\bf 90} 180602 (2003). 

\bibitem{Henk} H. van Beijeren, K. W. Kehr, and R. Kutner {\em Phys. Rev. B} {\bf 28}  5711 (1983). 

\bibitem{Roden} C. R\"odenbeck, J. K\"arger, and K. Hahn {\em Phys. Rev. E}
{\bf 57} 4382 (1998). 


\bibitem{Taloni} A. Taloni, and F. Marchesoni {\em Phys. Rev. Lett.} {\bf 96}, 020601 (2006).

\bibitem{Tobias} L. Lizana, T. Ambj\"ornsson 
{\em Phys. Rev. Lett.} {\bf 100}, 200601 (2008).

\bibitem{Aslangul} Cl. Aslangul {\em Europhysics Letters} {\bf 44} 284 (1998).   

\bibitem{Flomenbom} O. Flomenbom, and A. Taloni 
{\em Europhysics Letters} {\bf 83} 20004 (2008).

\bibitem{remark1}  Physically
this might be important for dynamics in slit pores,
where it was shown that non-interacting particles, exhibit super-diffusion, due to forward scattering from
the walls of pore \cite{Eok}. 

\bibitem{Risken}  H. Risken {\em The Fokker-Planck Equation} Springer-Verlag 
(Berlin) (1984). 

\bibitem{Jepsen} D. W. Jepsen {\em J. Math. Phys.} (N.Y) {\bf 6} 405 (1965). 

\bibitem{remark3} Levitt \cite{Levitt}
uses a Poissonian approximation to calculate
the motion of a tagged particle in the absence of a potential field.
We find that for our
aim, where an external force is acting on the particles,
the Poissonian approximation is wrong. This can be shown by an exact
calculation of the equilibrium distribution of the tagged particle,
using Boltzmann-Gibbs canonical ensemble. 

\bibitem{Montroll} E. W. Montroll and B. J. West in {\em Fluctuation Phenomena}
E. W. Montroll and J. L. Lebowitz Editors, Studies in Statistical Mechanics
Vol. VII North Holland, Amsterdam (1979).

\bibitem{remark4} $f(x_0)=0$ if $x_0<0$, $f(x_0)\ge 0$ and
$\int_0 ^\infty f(x_0) {\rm d} x_0 = 1$.



\bibitem{remark} To obtain Eq. (\ref{eq15cc}) we need the following integral
$ \int_0 ^\infty \exp( - \eta^2 y^2 / 2) \mbox{Erf} \left( y / \sqrt{2} \right) {\rm d} y =
\sqrt{2/ \pi} \mbox{arccot}( \eta ) / \eta$. 

\bibitem{remark5} We compared our analytical result with 
simulations of  \cite{Tobias} in Fig. 4. 


\bibitem{Band} 
T. Bandyopadhyay {\em Europhysics Letters} {\bf 81} 16003 (2008). 

\end{thebibliography}
\end{document}